\documentstyle[12pt]{article}
\newcommand{\be}{\begin{equation}}
\newcommand{\ee}{\end{equation}}
\newcommand{\ba}{\begin{eqnarray}}
\newcommand{\ea}{\end{eqnarray}}

\begin{document}
\begin{center}
{\large\bf EXACT SOLVABILITY OF TWO-DIMENSIONAL REAL SINGULAR MORSE POTENTIAL}\\

\vspace{0.3cm} {\large \bf M.V. Iof\/fe$^{1}$,
D.N. Nishnianidze$^{1,2}$
}\\
\vspace{0.2cm}
$^1$St.-Petersburg State University,
198504 St.-Petersburg, Russia\\
$^2$ Kutaisi State University, 4600 Kutaisi, Georgia\\
\today
\end{center}
\hspace*{0.5in}
\begin{minipage}{5.0in}
{\small The supersymmetric approach in the form of second order intertwining relations
is used to prove {\it the exact solvability} of two-dimensional
Schr\"odinger equation with generalized two-dimensional Morse potential for $a_0=-1/2$. This
two-parametric model is not amenable to conventional separation of variables, but it
is completely integrable: the symmetry operator of fourth order in momenta exists.
{\it All bound state} energies are found explicitly, and all corresponding wave functions
are built analytically. By means of
shape invariance property, the result is extended to the hierarchy of Morse models with arbitrary
integer and half-integer values $a_k=-(k+1)/2.$
\\
\vspace*{0.1cm} PACS numbers: 03.65.-w, 03.65.Fd, 11.30.Pb }
\end{minipage}
\vspace*{0.4cm}

{\it Introduction.} -- Very limited list of exactly solvable one-dimensional models is known up to now:
harmonic oscillator, singular
harmonic oscillator, Coulomb, Morse, P\"oschl-Teller potentials, and some others \cite{infeld}.
As for complete solution of the Schr\"odinger
equation in {\it two-dimensional space}, no regular approach is known. In practice,
only the conventional separation of variables \cite{miller}
provides the sole method to solve two-dimensional stationary quantum problems \cite{c}.
One of the fundamental problems of modern Quantum Mechanics is to enlarge as much as possible
the variety of exactly solvable multi-dimensional $(d\geq 2)$ systems.

The supersymmetric approach has provided a powerful impulse for
new development in analytical studies
both in one-dimensional  \cite{cooper} and multi-dimensional \cite{abei} Quantum Mechanics.
But to date, even among two-dimensional potentials, {\it not amenable} to
standard separation of variables \cite{david}, very few results on partially solvable models were obtained.
By means of supersymmetry, several two-dimensional models were found \cite{new} - \cite{innn},
for which {\it a part} of their
spectra and wave functions were found analytically (partial solvability).
It is necessary to stress that all constructed systems are
completely integrable, since the symmetry operators of fourth order in momenta were built for them.
Two specific supersymmetric techniques - the special SUSY-separation of variables and two-dimensional shape invariance -
allowed to investigate two-dimensional singular generalizations of Morse \cite{new}
and P\"oschl-Teller \cite{iv}
potentials. The first method explores the most important constituent of SUSY algebra - SUSY intertwining
relations. The second method is the natural generalization of the one-dimensional shape invariance approach \cite{genden},
which provided an elegant algebraic algorithm for solution of a class of Schr\"odinger equations. In one-dimension,
the class of shape invariant potentials coincides practically with the known variety of exactly solvable
models \cite{infeld}.

Among all possible two-dimensional intertwining relations with supercharges of second order in derivatives,
a subclass exists, where {\it one} of intertwined Hamiltonians {\it is amenable} to standard separation of variables
due to specific choice of parameters of the model. Such situation was studied already
for the {\it complex} two-dimensional Morse potential \cite{exact}, which is free of any singularities.
In the present Letter, we investigate the model, which is much more interesting physically -
the {\it real} singular two-dimensional Morse potential with parameter value $a_0=-1/2$.
Namely, we prove that, being completely integrable,
this two-parametric model is exactly solvable as well, i.e. all its eigenvalues and eigenfunctions are known analytically.
After that, we use two-dimensional shape invariance of the model to enlarge the class
of exactly solvable two-dimensional systems to a hierarchy of Morse potentials with parameter
values $a_k=-(k+1)/2;\,\, k=1,2,\ldots .$

\vspace*{0.3cm}
{\it Morse potential with $a_0=-1/2$.} -- Two-dimensional generalization of Morse potential \cite{new,ioffe}
is defined \cite{d} as:
\be
 \widetilde{V}(\vec x)=\frac{\alpha^2a(2a-1)}{\sinh^2(\frac{\alpha
 x_-}{2})}+U(x_1)+U(x_2), \,\, \vec x = (x_1, x_2),\,\, x_{\pm}\equiv x_1 \pm x_2,
 \label{morse}
\ee
where $a$ is an arbitrary real parameter, and $U(x_i)$ are one-dimensional Morse potentials:
\be
U(x_i)= A(e^{-2\alpha x_i}-2e^{-\alpha x_i});\,\,i=1,2 \quad A=Const >0; \,\,\alpha = Const >0,
\label{morse1}
\ee

The Hamiltonian $\widetilde H(\vec x)=-\Delta^{(2)}+\widetilde V(\vec x)$ with potential (\ref{morse})
and {\it the partner} Hamiltonian $H(\vec x)$ with potential:
\be
V(\vec x)=\frac{\alpha^2a(2a+1)}{\sinh^{2}(\frac{\alpha
 x_-}{2})}+U(x_1)+U(x_2).\label{2}
\ee
are intertwined \cite{david} - \cite{ioffe}
\be
\widetilde{H}(\vec x)Q^+=Q^+H(\vec x);\quad Q^-\widetilde H(\vec x)=H(\vec x)Q^-
\label{intertw}
\ee
by the second order supercharges:
\ba
Q^{\pm}&=&4\partial_+\partial_- \pm 4a\alpha\partial_- \pm 4a\alpha
\coth\frac{\alpha x_-}{2}\partial_+ + 4a^2\alpha^2\coth\frac{\alpha
x_-}{2}-\nonumber\\&-& A\biggl[e^{-2\alpha x_1}-2 e^{-\alpha x_1}
- e^{-2\alpha x_2}+2 e^{-\alpha x_2}\biggr];\quad \partial_{\pm}=
\frac{\partial}{\partial x_{\pm}}
\label{superch}
\ea

One can notice that $\widetilde V$ and $V$ differ from each other by a coefficient
in front of singular term only, this property being the origin of
shape invariance of the model \cite{shape}. By construction, each intertwined Hamiltonian obeys the symmetry operator
of fourth order in momenta \cite{new, ioffe},
which is expressed in terms of supercharges $Q^{\pm}.$ Specifically,
the operator $\widetilde R=Q^+Q^-$ commutes with Hamiltonian $\widetilde H(\vec x),$ and operator $R=Q^-Q^+$ - with
$H(\vec x).$

For the particular choice of parameter $a_0=-1/2,$ the partner Hamiltonian (\ref{2})
is simplified essentially - it allows the conventional separation of variables.
Moreover, each of one-dimensional problems, obtained after separation, is exactly solvable
one-dimensional Morse potential (\ref{morse1}). The discrete spectrum of this
model is well known \cite{landau}:
\be
\epsilon_n=-\alpha^2s_n^2;\quad
s_n\equiv\frac{\sqrt{A}}{\alpha}-n-\frac{1}{2} >0;\quad n=0,1,2,\ldots ,
\label{spectrum1}
\ee
and wave functions are expressed in terms of degenerate hypergeometric
functions \cite{bateman}:
\be
\eta_n(x_i) = \exp(-\frac{\xi_i}{2}) (\xi_i)^{s_n}
F(-n, 2s_n +1; \xi_i);\quad\,\,\xi_i\equiv \frac{2\sqrt{A}}{\alpha}\exp(-\alpha x_i).
\label{hyper}
\ee
Due to separation of variables, the quantum problem with the Hamiltonian $H(\vec x)$
from (\ref{2}) is exactly solvable. Its energy eigenvalues are:
\be
E_{n,m}=E_{m,n}=\epsilon_n+\epsilon_m,
\label{Enm}
\ee
being two-fold degenerate for $n\neq m.$ The corresponding eigenfunctions can be chosen as
symmetric or (for $n\neq m$) antisymmetric combinations:
\be
\Psi^{S(A)}_{E_{n,m}} = \eta_n(x_1)\eta_m(x_2)\pm\eta_m(x_1)\eta_n(x_2).
\label{psinm}
\ee

Our aim here is {\it to solve completely} the quantum problem for
$\widetilde H(\vec x)$ with potential (\ref{morse}) for $a_0=-1/2.$
The main tool are the SUSY intertwining relations (\ref{intertw}), providing \cite{ioffe}
connections between spectra and wave functions of partner Hamiltonians, which
are almost isospectral (up to zero modes and singular properties of $Q^{\pm}$).

In general, we {\it may expect} three kinds of levels of $\widetilde H(\vec x).$

(i). The levels, which coincide with (\ref{Enm}). Their wave functions can be
obtained from (\ref{psinm}) by means of intertwining relations (\ref{intertw}).

(ii). The levels, which were absent in the spectrum of $H(\vec x)$, if some wave functions
of $\widetilde H(\vec x)$ are simultaneously
the zero modes of the supercharge operator $Q^-.$ Then the second intertwining relation
in (\ref{intertw}) {\it would not give} any partner state among bound states of $H(\vec x).$

(iii). The levels, which were also absent in the spectrum of $H(\vec x)$, if some wave functions
of $\widetilde H(\vec x)$ become {\it nonnormalizable} functions after action of operators $Q^-.$

We will analyze these three classes of possible bound states of $\widetilde H(\vec x)$ one after another.

{\bf (i).} The first supersymmetric intertwining relation in (\ref{intertw}) gives immediately
the two-fold degenerate wave functions of $\widetilde H(\vec x)$
with energies (\ref{Enm}), as $\widetilde\Psi_{E_{nm}}=Q^+\Psi_{E_{nm}}.$
Using the explicit formulas (\ref{superch}) for the supercharge $Q^+$ and for
one-dimensional Schr\"odinger equation (\ref{morse1}), one can
rewrite the symmetric (antisymmetric) wave functions in the form:
\ba
\widetilde{\Psi}^S_{E_{n,m}} = Q^+\Psi^A_{E_{n,m}}=(\epsilon_m-\epsilon_n)
\Psi^S_{E_{n,m}}+D\Psi^A_{E_{n,m}},\label{6}\\
\widetilde{\Psi}^A_{E_{n,m}}\equiv Q^+\Psi^S_{E_{n,m}}=(\epsilon_m-\epsilon_n)
\Psi^A_{E_{n,m}}+D\Psi^S_{E_{n,m}},\label{7}
\ea
where the differential operator $D$ is defined as:
\be
D=\frac{\alpha^2}{\xi_2-\xi_1}[\xi_1+\xi_2+2\xi_1\xi_2
 (\partial_{\xi_1}+\partial_{\xi_2})].\label{8}
\ee

Since this operator is singular for $\xi_1=\xi_2,$ normalizability of
$\widetilde{\Psi}^{S,A}_{E_{n,m}}$ depends crucially on the behavior of
$\Psi^{A,S}_{E_{n,m}}$ on the line $\xi_1=\xi_2.$ The symmetric
wave functions on this line are
$\Psi^S_{E_{n,m}}(\xi_1,\xi_2=\xi_1) \sim \eta_n(\xi_1)\eta_m(\xi_1),$
and they have no zero multipliers $(\xi_1-\xi_2)$, which could compensate the
singularity of operator $D$. Therefore, the wave functions
$\widetilde{\Psi}^A_{E_{n,m}} = Q^+\Psi^S_{E_{n,m}}$
are certainly nonnormalizable, and they do not correspond to the physical bound states.

Vice versa, action of the operator $D$ in (\ref{6}) on the antisymmetric function
$\Psi^A_{E_{n,m}}$ leads to nonsingular result. To observe this property, one has
to use the explicit form (\ref{hyper}) of one-dimensional eigenfunctions.
After separation of exponential multiplier $\exp[-(\xi_1+\xi_2)/2]$,
the rest of function $\Psi^A_{E_{n,m}}$ is a polynomial
in $\xi_1, \xi_2.$ Due to antisymmetry, this polynomial vanishes for $\xi_1=\xi_2,$
since it contains the multiplier $(\xi_1-\xi_2),$ compensating the singularity in $D.$
Moreover, careful study shows that owing to the interplay between two terms in the r.h.s. of
(\ref{6}), the symmetric wave functions $\widetilde\Psi^S_{E_{n,m}}\sim
(\xi_1-\xi_2)^2$ at the singular point $\xi_1=\xi_2.$

It is possible to investigate efficiently the normalizability of wave functions
$\widetilde{\Psi}^S_{E_{n,m}}$ by the indirect algebraic method as well.
One can check straightforwardly that for $a_0=-1/2$ the operator $R=Q^-Q^+$ can be rewritten as:
\be
R=\biggl(h_1(x_1)-h_2(x_2)\biggr)^2 + 2\alpha^2\biggl(h_1(x_1)+h_2(x_2)\biggr)+\alpha^4,\label{9}
\ee
where $h_i(x_i)\equiv -\Delta_i + U(x_i)$ are one-dimensional Schr\"odinger
operators with Morse potential (\ref{morse1}). Acting by this operator on
the antisymmetric function $\Psi^A_{E_{n,m}},$ one obtains:
\ba
R\Psi^{A}_{E_{n,m}}(\vec x)&=&[(\epsilon_n-\epsilon_m)^2+2\alpha^2E_{n,m}+\alpha^4]
\Psi^{A}_{E_{n,m}}(\vec x)=\nonumber\\
&=&\alpha^4[(n-m)^2-1][(s_n+s_m)^2 - 1]\Psi^{A}_{E_{n,m}}(\vec x)\equiv
r_{n,m}\Psi^{A}_{E_{n,m}}(\vec x).\label{10}
\ea
This relation helps to find the norm of wave functions
$\widetilde{\Psi}^S_{E_{n,m}} :$
\be
\|\widetilde{\Psi}^{S}_{E_{n,m}}\|^2=
\langle\Psi^{A}_{E_{n,m}}\mid Q^-Q^+\mid \Psi^{A}_{E_{n,m}}\rangle=
r_{n,m}\|\Psi^{A}_{E_{n,m}}\|^2,
\label{11}
\ee
where we have used that $Q^{\pm}$ are mutually conjugate on the considered space. We stress
that off-integral terms in (\ref{11}) at the singular point $x_1=x_2$ disappear due
to vanishing behavior of normalizable wave functions $\widetilde\Psi^S(\vec x),$
mentioned above (see below Eqs.(\ref{44}),(\ref{55}) and nearby).

By definition, $\Psi^{A}_{E_{n,m}}$ in (\ref{11}) are zero
identically for $m=n.$ In its turn, the wave functions $\Psi^A_{E_{n,n\pm 1}}$ are annihilated
\cite{exact} by the action of $Q^+,$ since the corresponding $r_{n,m}$ in (\ref{10})
vanish for $m=n\pm 1.$
It follows from (\ref{10}), that the norms of all other $\widetilde{\Psi}^{S}_{E_{n,m}}$
(with $\mid n-m\mid\,\geq 2$) are finite and positive. Therefore, the first class of levels of $\widetilde H(\vec x)$
consists of nondegenerate energy levels (\ref{Enm}) with normalizable symmetric wave functions
$\widetilde\Psi^S_{E_{nm}}$ from (\ref{6}) with $\mid n-m\mid\,\geq 2.$

{\bf (ii).} This class of possible bound states of the Hamiltonian $\widetilde H(\vec x)$
with $a=-1/2$ consists of the
normalizable zero modes of the supercharge $Q^-$ from (\ref{superch}). In \cite{new} the variety
of normalizable zero modes of the Hermitian conjugated supercharge
$Q^+$ was studied in detail for arbitrary values of $a.$ In particular, it was shown that these
zero modes are normalizable, and no fall to the center occurs for the specific range for values
of $a.$
According to expressions (\ref{superch}), $Q^-$ can be obtained from $Q^+$ by replacing $a\to -a.$
Therefore, completely analogous analysis of zero modes of $Q^-$ will lead to the analogous, {\it but
positive}, interval for $a :$
\be
a \in (\frac{1}{4}+\frac{1}{4\sqrt{2}}\, ,\, +\infty ),
\label{interval}
\ee
which, for certain, does not contain the value $a_0=-1/2.$ This means that
no normalizable bound states of this class exist for $\widetilde H.$

{\bf (iii).} We have to study an opportunity that operator $Q^-$ destroys normalizability of some
eigenfunctions
of $\widetilde H.$ It could occur due to singular character
of $Q^-$ at $x_1 = x_2.$ It is convenient to choose the polar coordinates $\xi, \varphi$
in the vicinity of this line in the plane $(\xi_1, \xi_2)$:
\ba
\xi_1=\xi\cos\varphi;\,\,\, \xi_2=\xi\sin\varphi;\,\,\, \varphi \equiv\pi/4-\theta ;\,\,\theta\sim 0.
\nonumber
\ea

In these variables the potential $\widetilde V(\vec x)$ reads:
\ba
\widetilde{V}=
\alpha^2\biggr[\frac{\cos2\theta}{\sin^2\theta}+\frac{1}{4}\xi^2-
(s_1+\frac{3}{2})\sqrt{2}\xi\cos\theta\biggl]
\sim  \alpha^2 \frac{1}{\theta^2};\quad \theta\sim 0 \nonumber
\ea
Thus, the Hamiltonian $\widetilde{H}(\vec x)$ effectively acts as:
\be
\widetilde{H}=-\Delta^{(2)}+ \widetilde V(\vec x) \sim -\alpha^2\biggl[
\frac{1}{2}\partial^2_{\theta}
- \frac{1}{\theta^2}\biggr];\quad \theta\sim 0 ,\label{44}
\ee
and two kinds of behavior of its eigenfunctions are possible:
\be
\widetilde\Psi\sim \theta^2 \quad or\quad \widetilde\Psi\sim\theta^{-1}.\label{55}
\ee
Only the first one is normalizable at $\theta \sim 0$ with the measure
$dx_1dx_2=d\xi d\theta /\xi\cos 2\theta.$

In the same vicinity, the supercharge operator $Q^-$ acts as:
\ba
Q^-\sim \alpha^2 \biggl(\xi\partial_{\xi}-1\biggr)\biggl(\partial_{\theta}+\frac{1}{\theta}\biggr).
\nonumber
\ea
This operator is not able to transform normalizable wave function of (\ref{55})
to nonnormalizable. Therefore, the third class of possible wave functions
$\widetilde\Psi$ of $\widetilde H(\vec x)$ {\it does not exist} too.

Summing up the above analysis, we found that the spectrum of two-dimensional Hamiltonian
$\widetilde H(\vec x)$ with $a_0=-1/2,$ which is not amenable to separation of variables, consists only
of the bound states with energies (\ref{Enm}) for $|n-m|>1.$ This spectrum is bounded from above
by the condition of positivity of $s_n, s_m$ in (\ref{spectrum1}): $n,m < \sqrt{A}/\alpha - 1/2.$
The corresponding wave functions are
obtained analytically, according to (\ref{hyper}), (\ref{psinm}), and (\ref{6}).

\vspace*{0.3cm}
{\it Hierarchy of Morse potentials with $a_k=-(k+1)/2$.} -- In what follows we denote the exactly solvable
Hamiltonians $H(\vec x),\,\,\widetilde H(\vec x)$ investigated
above, as $H(\vec x; a_0),\,\,\widetilde H(\vec x; a_0).$ The motivation is that
we will consider the Hamiltonians
$H(\vec x; a_k),\, \widetilde H(\vec x; a_k)$ with arbitrary negative integer and half-integer
values $a=a_k=-(k+1)/2;\,\, k=0,1,2,\ldots .$
We will prove that all these Hamiltonians are also exactly solvable.

The important property is that due to identity $a_{k-1}(2a_{k-1}-1)=a_{k}(2a_{k}+1):$
\be
\widetilde{H}(\vec x; a_{k-1})=H(\vec x; a_{k});\,\, k=1,2,\ldots .
\label{kk}
\ee
This means that the following chain (hierarchy) of Hamiltonians can be built:
\be
H(\vec x; a_0)\div\widetilde{H}(\vec x; a_0)=H(\vec x; a_1)\div \widetilde{H}(\vec x; a_1)=
 \ldots \div\widetilde{H}(\vec x; a_{k-1})=H(\vec x; a_{k})\div \widetilde H(\vec x; a_{k}),
\label{chain}
\ee
where the sign $\div $ between two Hamiltonians denotes their intertwining by $Q^{\pm}(a_i).$

Since the Hamiltonian $\widetilde{H}(\vec x; a_0)$ (and therefore $H(\vec x; a_1)$) was shown
to be exactly solvable, we can use the intertwining relations from (\ref{chain}) to prove exact solvability of
$\widetilde H(\vec x; a_1)$ too. Considering this procedure for the general length of the chain (\ref{chain}),
we will illustrate results for the first section $k=1$ of the chain, with the
Hamiltonian under study $\widetilde H(\vec x; a_1=-1).$

It was shown above, that bound states of the Hamiltonian $\widetilde{H}(\vec x; a_0)=H(\vec x; a_1)$ are
described by the symmetric functions $\widetilde\Psi^S_{E_{m,n}}(\vec x; a_0)\equiv\Psi^S_{E_{m,n}}(\vec x; a_1)$
of (\ref{6}) and energy eigenvalues (\ref{Enm}) with $|n-m|>1.$ Therefore,
the functions
\be
\widetilde{\Psi}^A_{E_{n,m}}(\vec x; a_1)=Q^+(a_1)\Psi^S_{E_{n,m}}(\vec x; a_1)=
Q^+(a_1)Q^+(a_0)\Psi^A_{E_{n,m}}(\vec x; a_0);\,\, |n-m|>1
\label{333}
\ee
(if they are normalizable) are the wave functions of the partner Hamiltonian
$\widetilde H(\vec x; a_1).$ The only exclusion
concerns possible zero modes of $Q^+(a_1).$
It was shown in \cite{new}, that the wave functions of $H(\vec x; a_k)$
may coincide with zero modes of supercharge $Q^+(a_k)$ for energy levels:
\ba
E_{n,n+k+1}=-2\alpha^2(2a_ks_n+s_n^2)-4\alpha^2a_k^2=-\alpha^2(s^2_n+s^2_{n+k+1}).
\nonumber
\ea
In particular for $k=1,$ operator $Q^+(a_1),$ acting on the wave functions $\Psi_{E_{n,n\pm 2}}(\vec x; a_1)$
of $H(\vec x; a_1),$ may annihilate them, not admitting to the spectrum of $\widetilde H(\vec x; a_1).$

Analogously in the general case, up to zero modes of $Q^+$ the functions
\be
\widetilde{\Psi}_{E_{n,m}}(\vec x; a_{k})=Q^+(a_{k})\Psi_{E_{n,m}}(\vec x; a_{k})=
Q^+(a_{k})Q^+(a_{k-1})\ldots Q^+(a_0)\Psi^A_{E_{n,m}}(\vec x; a_0) \label{33333}
\ee
(again, if normalizable) are the wave functions of $\widetilde H(\vec x; a_k)$
with energies $E_{n,m}=-\alpha^2(s_n^2+s_m^2).$ The symmetries of wave functions (\ref{33333})
alternate and depend on the length of chain (\ref{chain}).

In order to control the situation with possible zero modes of $Q^+$ and normalizability of
the states (\ref{33333}),
we will use the following algebraic trick.
For general value of $a_k,$ the identity between Hamiltonians $H(\vec x; a_{k})=\widetilde{H}(\vec x; a_{k-1})$ leads
to identity (up to a function of the Hamiltonian itself) between their symmetry operators $R(a_{k})$
and $\widetilde R(a_{k-1}).$
 This relation can be derived straightforwardly by algebraic manipulations with result:
\be
Q^-(a_{k})Q^+(a_{k})=Q^+(a_{k-1})Q^-(a_{k-1})+\alpha^2(2k+1)
\biggl[2\widetilde{H}(\vec x; a_{k-1})+\alpha^2(2k^2+2k+1)\biggr];\label{444}
\ee
In the case $k=1$, Eq.(\ref{444}) can be used to transform the norm of the function (\ref{333}):
\ba
&&\|\widetilde{\Psi}_{E_{n,m}}(\vec x; a_1)\|^2=\langle\Psi^A_{E_{n,m}}(\vec x; a_0)|Q^-(a_0)Q^-(a_1)
Q^+(a_1)Q^+(a_0)|\Psi^A_{E_{n,m}}(\vec x; a_0)\rangle = r_{n,m} \label{555}\\
&& (r_{n,m}+6\alpha^2E_{n,m}+15\alpha^4)\|\Psi^A_{n,m}(a_0)\|^2
=\alpha^4r_{n,m}[(m-n)^2-4][(s_n+s_m)^2-4]\|\Psi^A_{n,m}(a_0)\|^2,\nonumber
\ea
where definitions (\ref{10}) for $r_{n,m}$ were explored. From Eq.(\ref{555}),
the (real) functions $\widetilde{\Psi}_{n,n\pm 2}(\vec x; a_1)$
vanish identically for $m=n\pm 2,$ being zero modes of
$Q^+(a_1).$ Therefore, energies $E_{n,n+2}$ are absent in the spectrum
of $\widetilde H(\vec x; a_1).$ Vise versa, due to the same Eq.(\ref{555}) the norms of all other
$\widetilde{\Psi}_{n,m}(\vec x; a_1)$ with $|n-m|>2$ are positive and finite.

This analysis can be generalized to {\it arbitrary} integer values of $k.$ The proof is performed
step by step, by using Eq.(\ref{444}) for indices $k, k-1,\ldots ,1,0$ in calculation of the norm:
\be
\|\widetilde{\Psi}_{E_{n,m}}(\vec x; a_k)\|^2=\langle\Psi^A_{E_{n,m}}(\vec x; a_0)|Q^-(a_0)\ldots Q^-(a_k)
Q^+(a_k)\ldots Q^+(a_0)|\Psi^A_{E_{n,m}}(\vec x; a_0)\rangle .
\label{555a}
\ee
After some algebra, one obtains that
\ba
\|\widetilde{\Psi}_{E_{n,m}}(\vec x; a_k)\|^2
&&=\langle\Psi^A_{E_{n,m}}(\vec x; a_0)|R(a_0)\biggl(R(a_0)+\Gamma_0\biggr)
\biggl(R(a_0)+\Gamma_0+\Gamma_1\biggr)\ldots\nonumber\\&&\ldots\biggl(R(a_0)+\Gamma_0+\ldots +\Gamma_{k-1}\biggr)
|\Psi^A_{E_{n,m}}(\vec x; a_0)\rangle ;
\label{555b}\\
\Gamma_l&&\equiv\alpha^2(2l+1)[2H(a_0)+\alpha^2(2l^2+2l+1)].\nonumber
\ea
Inside the matrix element (\ref{555b}), $R(a_0)$ can be replaced by its eigenvalues (\ref{10}), and sums
$(\Gamma_0+\Gamma_1+\ldots +\Gamma_i)$  by:
\ba
\sum_{l=1}^{l=i}\biggl[2\alpha^2(2l+1)E_{n,m}+\alpha^4\biggl((l+1)^4-l^4\biggr)\biggr]=
\alpha^4\biggl[-2(s_n^2+s_m^2)i(i+2)+(i+1)^4-1\biggr]. \nonumber
\ea
Thus, the norm (\ref{555a}) can be factorized:
\be
\|\widetilde{\Psi}_{E_{n,m}}(\vec x; a_k)\|^2=\alpha^{4k}\|\Psi^A_{E_{n,m}}(\vec x;a_0)\|^2 r_{n,m}
\prod_{i=1}^{k}[(n-m)^2-(i+1)^2][(s_n+s_m)^2-(i+1)^2].
\label{norms}
\ee
One can conclude that the norm vanishes for $|n-m|\leq (k+1),$ and it is
finite positive for other $n,m.$ This means that the spectrum of $\widetilde H(\vec x; a_k)$
{\it includes} all energy levels $E_{n,m}$ of $H(\vec x; a_0)$ with $|n-m|>(k+1)$ {\it only}. All others
disappear due to zero modes of $Q^+(a_i).$

Similarly to the analysis (ii) above, no normalizable wave functions of $\widetilde H(\vec x; a_k)$ can be
annihilated by the operator $Q^-(a_k),$ since the values $a_k=-(k+1)/2$ lie again outside of the interval
(\ref{interval}), permitted for
their normalizability. The class (iii) of possible bound states of $\widetilde H(\vec x; a_k)$ is also empty.
To prove this fact, one has to consider the behavior of operators $\widetilde H,\,\,Q^-$ and
normalizable wave functions $\widetilde\Psi$ at the point of singularity $\theta\sim 0 :$
\ba
&&\widetilde H(\vec x; a_k)\sim -\frac{\alpha^2}{2}\biggl[\partial^2_{\theta}-\frac{k+1}{\theta^2}\biggr];\quad
\widetilde\Psi_{E_{n,m}}(\vec x; a_k)\sim \theta^{k+2}\nonumber\\
&&Q^-(a_k)
\sim \alpha^2 \biggl(\xi\partial_{\xi}-(k+1)\biggr)\biggl(\partial_{\theta}+\frac{k+1}{\theta}\biggr).\nonumber
\ea
Similar to the case $k=0,$ operators $Q^-(a_k)$ can not destroy normalizability
of $\widetilde\Psi_{E_{n,m}}(\vec x; a_k).$

Summarizing the obtained results, the spectra of Hamiltonians $\widetilde H(\vec x; a_k)$ are not degenerate.
They consist of the bound states with energies $E_{n,m},$ given by (\ref{Enm}) with indices $|n-m|>k+2.$
Their wave functions $\widetilde\Psi_{E_{n,m}}(\vec x; a_k)$ are given by Eq.(\ref{33333}) in terms of degenerate
(confluent) hypergeometric functions. Like in the case $a_0=-1/2$, the discrete spectra are bounded
from above by the conditions (\ref{spectrum1}): $n,m < \sqrt{A}/\alpha - 1/2.$

\vspace*{0.2cm}
The work was partially supported by grants RFFI 06-01-00186-a,
RNP 2.1.1.1112.


\begin{thebibliography}{299}
\bibitem{c}
We must recall also the class of exactly solvable models of $N$ particles on a line,
so called Calogero and Calogero-like models \cite{calogero}. In a particular case of $N=3,$ such models
can be interpreted as one-particle models on the plane \cite{ic}.
\end{thebibliography}
\end{document}